\documentclass[12pt]{article}
\usepackage{amsmath}
\usepackage[dvips]{graphicx}
\usepackage{times}
\begin{document}
\begin{titlepage}
\title{Indications for the new unitarity regime in the extensive air showers measurements}
\author{S.M. Troshin,
 N.E. Tyurin\\[1ex]
\small  \it Institute for High Energy Physics,\\
\small  \it Protvino, Moscow Region, 142281, Russia}
\normalsize
\date{}
\maketitle

\begin{abstract}
We note that the new unitarity regime when scattering amplitude
goes  beyond the black disc limit (antishadowing)
could help in the explanation  of the regularities such as knee in the energy spectrum, existence of
penetrating and long-flying particles and other features observed in the measurements of the
extensive air showers which originate from cosmic particles interactions with the atmosphere.\\[2ex]
\end{abstract}
\end{titlepage}
\setcounter{page}{2}

The experimental and theoretical studies of cosmic rays are the
 important source of astrophysical information
(cf. e.g. \cite{der})
and  they simultaneously provide a window to the future results of accelerator
studies of hadron interaction mechanism at the LHC\footnote{It should be noted
 that the value of the total
cross--section extracted from cosmic rays measurements significantly depend on the
particular  model for elastic scattering, because measurements of the extensive air showers
 provide information on inelastic
scattering cross--section only\cite{engel}.}.

\begin{figure}[thb]
\begin{center}
\includegraphics[width=90mm]{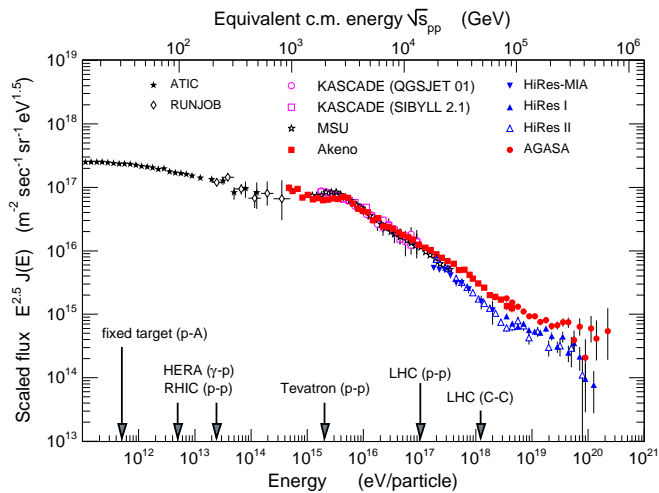}
\end{center}
\small{\caption{Scaled energy spectrum of the cosmic rays, figure from  \cite{engel}.}}
\label{ts:fig1}
\end{figure}
It can happen that the
investigations of cosmic rays will give us a clue that the hadron
interaction and mechanism of particle
generation is changing in the region of $\sqrt{s}=3-6$ TeV\cite{crrev,her}. Indeed, the
energy spectrum which follows simple power-like law $F(E)=cE^{-\gamma}$ changes
its slope in this energy region and becomes steeper: index $\gamma$ increases from
$2.7$ to $3.1$. It is important that the knee in the energy spectrum appears in the same
energy region where the penetrating and long--flying particles also start to appear in
the extensive air showers (EAS):
the absorbtion length is also changing from $\lambda=90$ $g/cm^2$ to
$\lambda=150$ $g/cm^2$ (cf. \cite{crrev}). There is also specific feature of the
events at the energies beyond knee such as alignment {cf. \cite{cask} and the
 references for the earlier papers therein}.
The above phenomena  were interpreted as a result of appearance
 of the new particles which have a small inelastic cross--section and/or
small inelasticity. These new particles can be associated with a manifestation of the
supersymmetry, quark--gluon plasma formation and other new mechanisms. However, there
is another possibility to treat those cosmic rays phenomena observed in EAS as the manifestations
 of the new unitarity regime
(antishadow scattering mode)
at such energies \cite{ashd}.

Unitarity of the scattering matrix $SS^+=1$ implies, in principle, an
existence at high energies $s>s_0$, where $s_0$ is a threshold\footnote{Model
estimates show that
new scattering mode starts to develop right beyond Tevatron energies, i.e. at
$\sqrt{s_0}\simeq 2$ TeV \cite{s0}, which corresponds to the energy in the laboratory system
$E\simeq 2$ $PeV$.}
 of the new scattering mode ---
antishadow one. It has been revealed in \cite{ashd} and
 described in some detail (cf.
\cite{echn} and references therein) and the most important feature
of this mode is the self-damping of the
inelastic channels contributions at small values of impact parameter --- antishadowing.
The antishadowing leads to $P(s,b=0)\to 1$ at $s\to\infty$, where $P$ is
a probability of the absence of the inelastic interactions,
$P(s,b)\equiv |S(s,b)|^2$, where $S$ is the elastic scattering
$S$--matrix.

Self-damping of the inelastic channels  leads to
 asymptotically dominating role of elastic scattering.
The cross--section of inelastic processes rises with energy as $\ln s$, while
  elastic and total cross--sections behave asymptotically as $\ln^2 s$.
The antishadow scattering mode could definitely be
observed at  the LHC energies and studies of the extensive air showers
originated from the cosmic particles interactions
with the atmosphere  provide evidence for it as we will argue in what follows.
Starting at some threshold energy $s_0$ (where amplitude reaches the black disk
 limit at $b=0$), antishadowing can occur at higher energies
 in the limited region of impact parameters $b<R(s)$ (while
 at large impact parameters only shadow scattering mode can be
 realized).

The inelastic overlap function $\eta(s,b)$ becomes peripheral
 when energy goes beyond $s=s_0$ (Fig.2).
At such energies the inelastic overlap function reaches its maximum
 value at $b=R(s)$, where $R(s)$ is the interaction radius,
 while the elastic scattering occurs at smaller values
of impact parameter, i.e.
$\langle b^2 \rangle_{el}<\langle b^2 \rangle_{inel}$. Note that
\[
\langle b^2 \rangle_{i}
\equiv \frac{1}{\sigma_i}\int_0^\infty b^2 \frac{d\sigma_i}{db^2},
\]
 where $i=tot,el,inel$
and
\[
\mbox{Im} f(s,b)\equiv \frac{1}{4\pi}\frac{d\sigma_{tot}}{db^2};\,\,
|f(s,b)|^2\equiv \frac{1}{4\pi}\frac{d\sigma_{el}}{db^2};\,\,
\eta(s,b)\equiv \frac{1}{4\pi}\frac{d\sigma_{inel}}{db^2}
\]
and unitarity condition in the impact parameter space is the following
\[
\mbox{Im} f(s,b)=|f(s,b)|^2+\eta(s,b),
\]
where  $f(s,b)$ is the elastic scattering amplitude.
The quantity $\langle b^2 \rangle$ is a measure of the reaction peripherality.
 Despite that the asymptotics for
$\sigma_{el}$ and $\sigma_{inel}$ are different, the quantities $\langle b^2 \rangle_{el}$ and $
\langle b^2 \rangle_{inel}$  have the same asymptotical energy dependence, proportional to $\ln^2 s$.

So, beyond the transition energy range there are two regions in impact
 parameter space: the central region where self-damping of inelastic channels occurs
(antishadow scattering at $b< R(s)$) and the peripheral region of
shadow scattering at $b> R(s)$.
\begin{center}
\begin{figure}[hbt]
\begin{center}
\includegraphics[width=90mm]{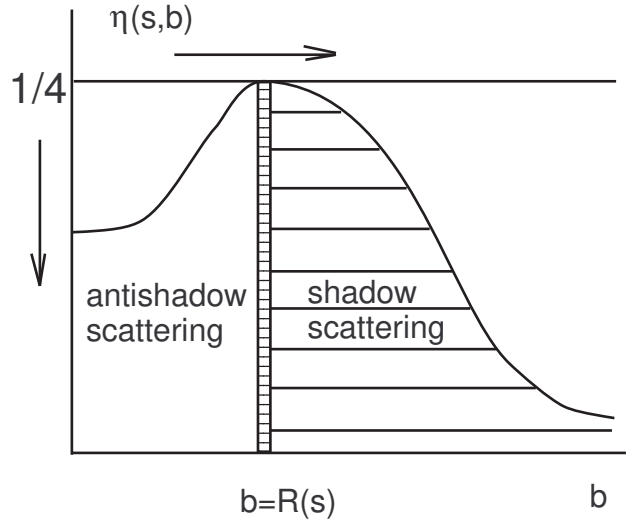}
\end{center}
{\small\caption{Impact parameter dependence of the inelastic overlap
function  in the framework of the unitarization scheme
 with antishadowing. Arrows indicate the directions of movement of minimum at $b=0$ and maximum at
 $b=R(s)$ with
 the energy increase. In the region of $b=R(s)$ the complete absorbtion takes place, i.e.
 $|S(s,b=R(s))|^2=0$.}}
 \end{figure}
\end{center}
At the energies ${s}\gg {s_0}$ small impact parameter scattering
is almost elastic one.

Thus head--on colliding particles  will provide appearance
of penetrating long-flying component in the  EAS
 and such particles will  spend only
small part  of
their energy for the production of secondaries. The head-on collisions
will lead to smaller  number of secondary particle and it will provide
 faster decrease of the energy spectrum of cosmic rays, i.e. it
  will result in the appearance of the knee. This qualitative picture
  will be explained in more detail  in what follows. It should be noted
  that this effect has a threshold in the energy dependence. It is also
  important to note that due to small probability of the sequential head-on
  collisions the number of events with penetrating particles also should be small.
  Nontheless, such events have been observed in the experiments
   PAMIR \cite{ari}.

Antishadowing leads to suppression of particle
production at small impact parameters:
\begin{equation}\label{mm}
\bar n(s)= \frac{1}{\sigma_{inel}(s)}{\int_0^\infty  \bar n
(s,b)\frac{d\sigma_{inel}}{db^2}db^2},
\end{equation}
i.e. multiplicity distribution
\[
P_n(s,b)\equiv \frac{1}{\sigma_{inel}(s)}\frac{d\sigma_{n}(s)}{db^2}
\]
 and mean multiplicity
$\bar n(s,b)$ in the impact parameter representation have
no absorptive corrections, but peripherality of $ {d\sigma_{inel}}/{db^2}$
 leads to suppression of particle
production at small impact parameters and the main contribution to
the integral  multiplicity $\bar n(s)$ comes from
the region of $b\sim R(s)$ (Eq. (\ref{mm})). This would lead to the events with
 alignment observed in EAS  and also to the imbalance
 between orbital angular momentum in the initial and final states since particles
 in the final state will carry  out large orbital
 angular momentum.
To compensate this orbital momentum spins of secondary particles should  become
  lined up, i.e. the spins of the produced particles should demonstrate
   significant
  correlations when the antishadow scattering mode appears \cite{sc}. Thus, the
   observed phenomena of
  alignment in EAS \cite{cask}
  and predicted spin correlations of final particles should have a common origin.
  The model estimate for the primary energy when these phenomena should appear is
  $E_0 \simeq 2$ $PeV$  --- $E_0$ is the energy
  when the new unitarity regime starts to develop at small impact parameters.

The detected particle composition of the EAS is closely related to the quantity known as
gap survival probability.
Antishadowing  leads to the nonmonotonous energy dependence of this
quantity \cite{gsp}. The gap survival probability, namely the probability to keep away
inelastic interactions  which can result in filling up by hadrons
the large rapidity gaps, reaches its
minimal values at the Tevatron highest energy and this is due to the fact that
 the scattering at this energy is very close to
the black disk limit at $b=0$ (Fig. 3).
\begin{figure}[thb]
\begin{center}
\includegraphics[width=80mm]{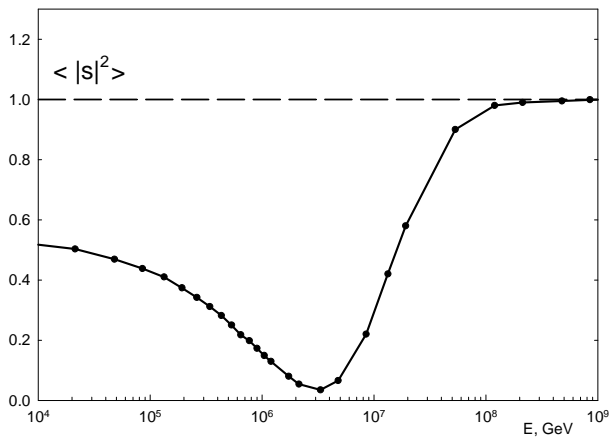}
\end{center}
\small{\caption{Energy dependence of gap survival probability.}}
\label{fig4}
\end{figure}
 It is clear that its higher value
means higher fraction for diffractive  component and consequently
the increasing of this component would result  in the enhancement of
the relative fraction
of protons in  the observed cosmic rays spectrum. Otherwise, decreasing of this quantity
will lead to increase of pionization  component and consequently to the increasing
number of muons
  observed as  multi-muon events.
Experiment reveals that relative fraction of protons in cosmic rays also shows
nonmonotonous energy dependence (cf. Fig. 4). To explain such dependence an additional
component is introduced {\it ad hoc} at the energies above $3\cdot 10^7$ GeV. It was shown
that account
of the antishadowing makes an introduction of this {\it ad hoc} component unnecessary.

The  inelasticity parameter $K$,
which is defined as ratio of the energy going to inelastic processes to the total energy, is
 important for the interpretation
of the EAS cascades developments. Its energy dependence is not
 clear and number of models predict the decreasing energy dependence while other
 models insist on the increasing energy behaviour at high energies
 (cf. e.g. \cite{shabel}). Adopting simple ansatz of geometrical models where parameter
 of inelasticity
 is related to inelastic overlap function we can use the following equation for
 $\langle K \rangle$ \cite{dias}
 \[
 \langle K \rangle=4\frac{\sigma_{el}}{\sigma_{tot}}
 \left(1-\frac{\sigma_{el}}{\sigma_{tot}}\right)
\]
to get a qualitative knowledge on the inelasticity energy dependence.
The estimation of inelasticity  based on the particular model with
 antishadowing \cite{s0} leads to increasing dependence of inelasticity with energy
 till $E\simeq 4\cdot 10^7$ GeV. In this region inelasticity reaches maximum value
 $\langle K \rangle = 1$, since ${\sigma_{el}}/{\sigma_{tot}}=1/2$ and then
 starts  to decrease at the energies where this ratio goes beyond the black
  disk limit $1/2$.
Such qualitative
 nonmonotonous energy dependence of inelasticity is the result of transition to
  the antishadowing
 scattering regime. The distribution on the inelasticity is related
to the distribution
 on the effective mass number, i.e. changes of $A$ are equivalent
 to changes of $\langle K \rangle$, and, for example,
  high-inelasticity primary proton interaction
 produces the same result at the ground level as the low-inelasticity primary interaction
 of the heavy nuclei \cite{jones}. The available experimental data on the average logarithm
 of the effective nuclear mass number, extracted from  the energy dependence of the depth
 of EAS maximum,  have large error bars, but they also
 indicate a nonmonotonous energy dependence with the maximum in the region
  $E\simeq (4-5)\cdot 10^7$ GeV
  \cite{hern}.

 It is also worth to note  that the maximum in inelasticity energy
 dependence, when the pionization component is maximal, is
 correlated with the minimum of the relative component of protons in the EAS,
 the following simple relation can be supposed
\[
{\Phi_p}/{\Phi_{all}}\sim 1-\langle K \rangle
\]
i.e. the relative
 proton component in the detected EAS should have  a non-monotonic energy dependence and this
is in agreement with the experimental analysis represented in Fig.~4.
\begin{figure}[thb]
\begin{center}
\includegraphics[width=90mm]{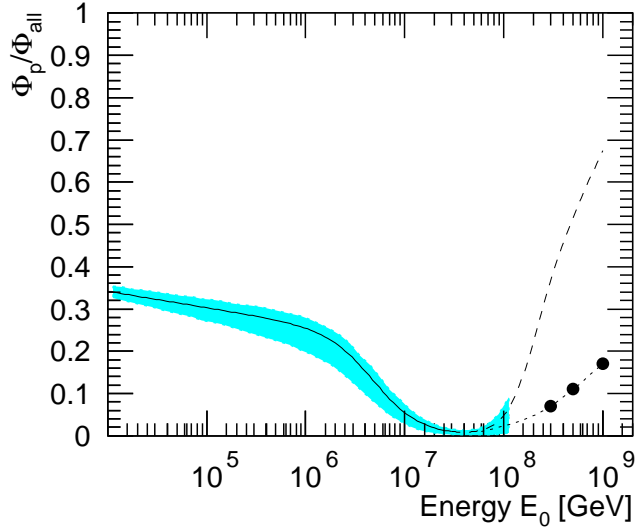}
\end{center}
\small{\caption{Relative fraction of protons in EAS,
figure is taken from  \cite{hern}.}}
\label{pfrac}
\end{figure}

 It should be noted that the behaviour of the ratio
  ${\sigma_{el}}/{\sigma_{tot}}$ when it goes to unity at
  $s\to\infty$ does not imply decreasing energy dependence of $\sigma_{inel}$.
  The inelastic cross--section $\sigma_{inel}$ increases monotonically
  and it grows as $\ln s$ at $s\to\infty$. Such a dependence of $\sigma_{inel}$ is
  in good agreement with the experimental data and, in particular, with the observed
   falling slope of the depth of shower maximum distribution \cite{gaisser}.
   The predicted numerical value of the inelastic cross-section is
   $\sigma_{inel}(s)\simeq 76$ $mb$ at the LHC energy $\sqrt{s}=14$ $TeV$. This value
   is also in a good agreement with the value for this quantity extracted from
    the proton-air inelastic cross-section \cite{hern}. This approach  provides
    a reasonable description \cite{jpg} of
the energy dependence of mean multiplicity and leads to
its power-like growth with a small exponent.

The relation of the knee and other effects observed in the EAS measurements
 with the
modification of particle generation mechanism is under discussion since the time when
they were discovered.
We propose here one particular realization of this idea --- an approach where
the corresponding particle generation mechanism in EAS is strongly affected by the unitarity effects and
 the energy region between the knee and the ankle coincides with
 the transition region to the  scattering mode where antishadowing develops at small and then at
 moderate values
 of impact parameter, i.e. the  energy spectrum of the primary cosmic particles
  $F_0(E)$ is modulated  by the significant variation of the scattering matrix $S$ in the energy region
starting from about $E_1\simeq 10^6$ GeV and finishing at about $E_2\simeq 10^9$ GeV and this
 resulting in the  regularities in the observed
spectrum $F(E)$ measured in the EAS studies. Below the energy $E_1$ and beyond the energy $E_2$ variation
of scattering matrix is slow and the primary energy spectrum $F_0$ is almost not affected.
It seems to be a rather natural explanation
of the observed  regularities in the EAS measurements and has a close interrelation
with the nonmonotonous energy dependence of gap survival probability and inelasticity.
This hypothesis is based on the saturation of the unitarity and can be experimentally
checked at the LHC \cite{echn}. The studies of the proton scattering in the forward
 region at the LHC will be very helpful for improving the
 interpretation of the results of the cosmic rays experiments.

\section*{Acknowledgement}
We are grateful to V.A. Petrov for the interesting discussions on the impact parameter dependence
of the mean multiplicity.


\small
\begin{thebibliography}{99}
\bibitem{der}
A. De R\'ujula,  Nucl. Phys. Proc. Suppl. 151, 23, 2006, [hep-ph/0412094, astro-ph/0411763].
\bibitem{engel}
R. Engel, T.K. Gaisser, P. Lipari, T. Stanev,
Phys. Rev. D 58,  014019, 1998;\\
R. Engel, astro-ph/0504358.
\bibitem{crrev}
T. Stanev, astro-ph/0411113;\\  S.I. Nikolsky,  V.G. Sinitsina,
Phys. Atom. Nucl. 67,  1900, 2004, ;\\  A.A. Petrukhin,
Proc. of the 28th International Cosmic Ray Conferences ICRC 2003, Tsukuba,
Japan, 31 Jul - 7 Aug 2003, 275.
\bibitem{her}
 J.R. H\"{o}randel, talk at
19th European Cosmic Ray Symposium, Florence, Italy, 30 Aug - 3 Sep 2004;
astro-ph/0501251.
\bibitem{cask}
T. Antoni et al., Phys. Rev. D 71, 072002, 2005.
\bibitem{ashd}
S. M. Troshin, N. E. Tyurin, Phys. Lett. B 316,  175, 1993.
\bibitem{hern}
J.R. H\"{o}randel, J. Phys. G  29,  2439, 2003.
\bibitem{echn}
S.M. Troshin, N.E. Tyurin,
Phys. Part. Nucl. 35,  555, 2004.
\bibitem{s0}
S.M. Troshin, N.E. Tyurin,
Eur. Phys. J. C  21,  679, 2001;\\
V.A. Petrov, A.V. Prokudin, S.M. Troshin, N.E. Tyurin,
J. Phys. G 27,  2225, 2001.
\bibitem{ari}
T. Arisawa, et al., Nucl. Phys. B 424,  241, 1994.
\bibitem{sc}
S.M. Troshin, Phys. Lett. B 597,  391, 2004.
\bibitem{gsp}
S.M. Troshin, N.E. Tyurin,
Eur. Phys. J. C  39,  435, 2005.
\bibitem{shabel}
Yu.M. Shabelski, R.M. Weiner, G. Wilk, Z. Wlodarczyk, J. Phys. G 18, 1281, 1992.
\bibitem{dias}
J. Dias de Deus, Phys. Rev. D 32,  2334, 1985;\\
S. Barshay, Y. Ciba, Phys. Lett. B 167,  449, 1985.
\bibitem{jones}
L.W. Jones, Nucl. Phys. B75A, 54, 1999.
\bibitem{gaisser} T.K. Gaisser et al., Phys. Rev. D 36, 1350, 1993.
\bibitem{jpg}
S.M. Troshin, N.E. Tyurin,
J. Phys. G 29,  1061, 2003.
\end{thebibliography}
\end{document}